\documentclass[superscriptaddress,preprint]{revtex4}
\usepackage{mathrsfs}
\usepackage{amssymb}
\usepackage[tbtags]{amsmath}
\usepackage{graphicx}
\usepackage{epsfig,graphicx,times}
\usepackage{color}
\usepackage{CJK}

\begin{document}
\begin{CJK*}{GBK}{song}
\title{Vacuum induced Berry phases in single-mode Jaynes-Cummings models}

\author{Yu Liu}
\affiliation{Quantum Optelectronics Laboratory, School of Physics
and Technology, Southwest Jiaotong University, Chengdu 610031,
China}
\author{L. F. Wei\footnote{ weilianfu@gmail.com}}
\affiliation{Quantum Optelectronics Laboratory, School of Physics
and Technology, Southwest Jiaotong University, Chengdu 610031,
China}
\author{W. Z. Jia}
\affiliation{Quantum Optelectronics Laboratory, School of Physics
and Technology, Southwest Jiaotong University, Chengdu 610031,
China}
\author{J. Q. Liang}
\affiliation{Institute of Theoretical Physics and Department of
Physics, Shanxi University, Taiyuan 030006, China}

\date{\today}
\begin{abstract}
Motivated by the work [Phys. Rev. Lett. 89, 220404 (2002)] for
detecting the vacuum-induced Berry phases with two-mode
Jaynes-Cummings models (JCMs), we show here that, for a
parameter-dependent single-mode JCM, certain atom-field states also
acquire the photon-number-dependent Berry phases after the parameter
slowly changed and eventually returned to its initial value. This
geometric effect related to the field quantization still exists,
even the filed is kept in its vacuum state. Specifically, a feasible
Ramsey interference experiment with cavity quantum electrodynamics
(QED) system is designed to detect the vacuum-induced Berry phase.

\end{abstract}
\maketitle

In 1984, Berry showed that the state of a quantum system can acquire
a purely geometric phase (called now as Berry phase), in addition to
the usual dynamical phase, after slowly changed and eventually
returned to its initial form~\cite{1}. Basically, Berry phase does
not depend on the dynamical properties of the system, but just
depends on the topological feature of the parameter space of the
evolved system. Up to now, Berry Phase has been found in various
systems, such as spins, polarized lights, atoms and so on~\cite{2}.
Also, recent studies have shown that the geometric phases can be
utilized to implement quantum logic gates for realizing quantum
computation~\cite{3,4,5}.

Quantized optical fields, as well as their interactions with atoms,
are the main objects in quantum optics. Originally, the famous
Jaynes-Cummings model (JCM)~\cite{6} is introduced to describe the
interaction between an undamped two-level atom and a non-decaying
single-mode quantized field, under the rotating-wave approximation.
This model has been widely generalized to treat various interactions
between atoms and photons. These include, e.g., multilevel atoms
interact with multimode quantized fields, and various multiphoton
processes in quantum optics~\cite{7}.
One of the basic phenomena in quantum optics is that, the vacuum of
a quantized field can behave as a physical reality with certain
observable effects. For example, in terms of vacuum fluctuations of
the quantized electromagnetic field~\cite{8} certain important
quantum effects, such as Lamb shifts and spontaneous emissions can
be well explained. Recent works~\cite{9,10} indicated that, the
vacuum of quantized optical field could also induce the observable
Berry phases. In order to observe these vacuum-induced geometric
effects, two filed modes were introduced~\cite{9,10} to interact
with a two-level atom. As a consequence, the experimental tests are
relatively complicated.

In this work, we show that only one field mode interacting with a
two-level atom could be utilized to detect the vacuum-induced Berry
phase. Beginning with a generic model, i.e., the $m$-quantum JCM, we
show how the desirable Berry phase can be acquired by an evolved
quantum state in a parameter-dependent single-mode JCM. Furthermore,
we design a Ramsey interference device involved with only one filed
mode to detect such a geometric effect related to the field
quantization.

The Hamiltonian of a $m$-quantum JCM~\cite{11}, i.e., a two-level
system coupled to a quantized mode via a $m$-photon process, can be
expressed as (under the rotating-wave approximation)
\begin{equation}
H=\nu a^{\dag}a +
\frac{\omega}{2}\sigma_{z}+\lambda_{m}(\sigma_{+}a^{m}+\sigma_{-}a^{\dagger
m}).
\end{equation}
Here, $a^{\dagger}$ and $a$ are the creation and annihilation
operators of the cavity field with frequency $\nu$. $\sigma _{+}$,
$\sigma _{-}$, and $\sigma _{z}$ are the Pauli operators of the
atom. $\omega $ is the transition frequency of the atom between the
excited state $|2\rangle $ and ground state $|1\rangle $, and
$\lambda_{m}$ the coupling coefficient between the atom and cavity
mode.
Under a time-dependent unitary transformation
$\hat{S}(t)=\exp(i\Delta_{m}\sigma_{z}t/2)$, the above Hamiltonian
can be rewritten as
\begin{equation}
H_{m}=\frac{\Delta_{m}}{2}\sigma_{z}+\lambda_{m}(\sigma_{+}a^{m}+\sigma_{-}a^{\dag
m})
\end{equation}
where $\Delta_{m}=\omega-m\upsilon$ is the detuning. The eigenstates
of such a Hamiltonian read
\begin{equation}
|\Psi _{n}^{-}\rangle =\cos \frac{\theta _{nm}}{2}|1,n+m\rangle
-\sin \frac{\theta _{nm}}{2}|2,n\rangle,
\end{equation}
and
\begin{equation}
|\Psi _{n}^{+}\rangle =\cos \frac{\theta _{nm}}{2}|2,n\rangle +\sin
\frac{\theta _{nm}}{2}|1,n+m\rangle,
\end{equation}
respectively. Above, $\{|n\rangle,\,n=0,1,2,...\}$ are the number
states the quantized bonsic field, and
$\theta_{nm}=\arccos[\,\,\Delta_{m} /\sqrt{\Delta_{m}
^{2}+4\lambda_{m} ^{2}(n+m)!/n!}\,\,]$.

Following Fuentes-Guridi {\it et. al.}~\cite{9}, we introduce a
phase-shift operation $U[\phi(t)]=\exp[-i\phi(t)a^{\dag}a]$ to
change the Hamiltonian (2) to the following parameter-dependent form
\begin{equation}
H_m(\phi)=\frac{\Delta_{m}}{2}\sigma_{z}+\lambda_{m}(\sigma_{+}a^{m}e^{im\phi}+\sigma_{-}a^{\dag
m}e^{-im\phi}).
\end{equation}
Obviously, such a $\phi$-dependent Hamiltonian describes a two-level
atom interacting (via a $m$-photon process) with a quantized field
mode. Here, the phase parameter $\phi(t)$ changes with the time $t$
and can be changed slowly from $0$ to $2\pi$ generating a cyclic
path in the parameter space during the evolution. As a consequence,
if the system begins with one of its eigenstates,
$|\Psi_{n}^{+}\rangle$ or $|\Psi_{n}^{-}\rangle$, then it returns to
such state but acquires a geometric phase (besides the dynamical one
not shown here)
\begin{eqnarray}
\gamma_{+}=i\int_{c}d\phi\langle\Psi_{n}^{+}|U^{\dag}(\phi)\frac{d}{d\phi}U(\phi)|\Psi_{n}^{+}\rangle
=m\pi ( 1-\cos \theta _{nm} )+2\pi n,
\end{eqnarray}
or
\begin{eqnarray}
\gamma_{-}=i\int_{c}d\phi\langle\Psi_{n}^{-}|U^{\dag}(\phi)\frac{d}{d\phi}U(\phi)|\Psi_{n}^{-}\rangle
=- m\pi ( 1-\cos \theta _{nm} )+2\pi (n+m).
\end{eqnarray}

It is seen that the geometric phase acquired here depends on the
photon number $n$. Physically, this nontrivial quantum effect can be
measured by using an interference procedure between the eigenstate
$|\Psi _{0}^{+}\rangle$ (or $|\Psi _{0}^{+}\rangle$) and the ground
state $|1,0\rangle$, for which no geometric phase is acquired.
Typically, if the system begins with the state $|2,0\rangle =\cos
\theta _{0m}/2|\Psi _{0}^{+}\rangle -\sin \theta _{0m}/2|\Psi
_{0}^{-}\rangle $, i.e., the field is in a vacuum state, the above
adiabatic operation performed on the degrees of freedom of the field
still yields a Berry phase
\begin{equation}
\gamma_{0m} =m\frac{\pi}{2} ( 1-\cos 2\theta _{0m}
)=m\frac{\Omega_{m}}{4},
\end{equation}%
\begin{figure}[tbp]
\includegraphics[width=12cm,height=6.6cm]{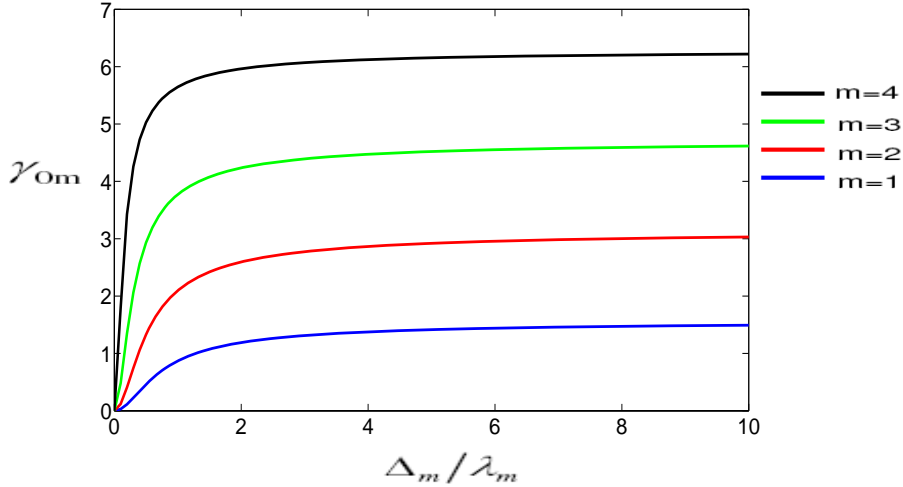}\newline
\caption{The blue (red, green, black) curve represents
 the vacuum induced Berry phase of $m=1 (m=2, m=3, m=4)$, respectively.}
\end{figure}
with the solid angle $\Omega_{m}=2\pi[1-\cos(2\theta_{0m})]$.
Fig.~1 shows how the above vacuum-induced Berry phase varies with
the parameter $\bigtriangleup_{m}/\lambda_m$. Through adiabatic
evolution, the initial state $|2\rangle$ coupling to vacuum mode in
cavity acquires a geometric phase. In this case, the atom-field
entanglement in the eigenstates (3), (4) cannot be
neglected~\cite{9}. Note that the expression (8) cannot only be
interpreted as a geometric phase of the two level system, as the
origin of the geometric phase is related to the vacuum fluctuation
of the field. Clearly, for a common
$\bigtriangleup_{m}/\lambda_m\equiv\Delta/\lambda$ the more quantum
$m$ corresponds to the greater vacuum-induced Berry phase.
Basically, the photon-dependent Berry phase shown in Eqs.~(6-7) is
due to the performance of the field quantization. Thus, even the
photon number of the field is $0$, the geometric phase is still
nontrivially induced. Any classical correspondence of such a
phenomenon does not exist.

Berry phases related to field quantization could be measured with
the usual one-photon JCMs, which had been experimentally
demonstrated in the well-known cavity QED systems. Indeed, various
quantum natures~\cite{12} of the radiation field interacting with
atoms have been successfully demonstrated with these systems.
Typically, a cavity QED experiment~\cite{10} involved with {\it two}
quantized bosonic modes was proposed to test the geometric phases
generated in a two-mode JCM. Below, we show that a cavity QED system
involved only one quantized bosonic mode could also be utilized to
test the above vacuum-induced Berry phase. Our proposed setup for
such a test is shown in Fig.~2, wherein an atom is emitted from the
source $O$ and then flies sequentially
across $R_1$, $C$ and $R_2$, and is finally detected in $I$.%
%
\par\begin{figure}[tbp]
\includegraphics[width=15.5cm,height=6cm]{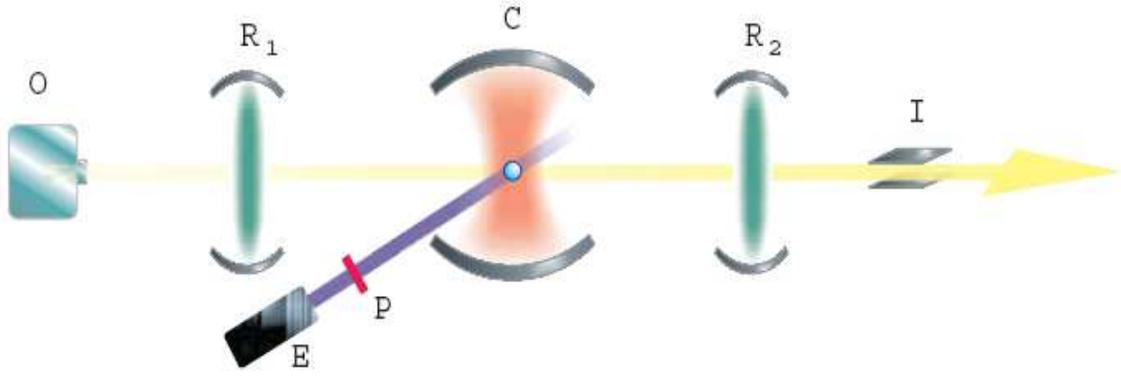}\newline
\caption{A experimental Ramsey interference setup~\cite{10} for
observing the Berry phase generated in the one-mode JCM. Here, an
atom is emitted from the source $O$, and flies sequentially across
the first Ramsey zone $R_1$, high-Q quantized cavity $C$ (wherein
the cavity is kept in the vacuum state and the desirable
vacuum-induced Berry phase is generated by a classical driving from
$E$), the second Ramsey zone $R_2$, and then is finally detected in
$I$. The information of the vacuum-induced Berry phase is extracted
by the measured atomic probability.}
\end{figure}

Initially, the atom is assumed to be prepared in the upper level
$|2\rangle$ in the source $O$ and then emitted. After the first
Ramsey zone $R_1$, the state of the atom reads
\begin{equation}
|\Psi_1\rangle=\cos(\frac{\Omega_{R1}\tau_1}{2})
|2\rangle+i\sin(\frac{\Omega_{R1}\tau_1}{2}) |1\rangle,
\end{equation}
with $\tau_1$ being the time spent by the atom inside the zone
$R_{1}$.

During the atom flies across the high-Q quantized cavity $C$, the
parameter-dependent Hamiltonian (5) can be obtained. For example, a
Raman configuration shown in Fig.~3 is utilized to achieve the
$\phi$-dependent one-photon JCM. Here, an auxiliary external
classical laser beam $E(t)$ is applied to drive the transition
$|2\rangle\leftrightarrow|3\rangle$ with the Rabi frequency
$\Omega_{L}=\Omega_0\exp(i\phi)$, while the quantized cavity mode
($a,\,a^\dagger$) couples to the transition
$|1\rangle\leftrightarrow|3\rangle$ with the strength $g$. The
Hamiltonian describing such a configuration in the interaction
picture reads ($\hbar\equiv 1$) (see, e.g.,~\cite{13})
\begin{equation}
H_{\rm int}=\Omega_{L}\sigma_{32}e^{-i\delta t}+g\sigma_{31}a
e^{-i\delta t}+H.c.,
\end{equation}
with $\delta$ being the detuning.
Generally, the corresponding time-evolution operator can be formally
expressed as
\begin{equation}
U_{I}(t)=1-i\int_{0}^{t}dt^{^{\prime }}H_{I}(t^{^{\prime
}})-\int_{0}^{t}dt^{^{\prime }}H_{I}(t^{^{\prime }})
\int_{0}^{t}dt^{^{\prime \prime }}H_{I}(t^{^{\prime \prime
}})+\ldots.
\end{equation}

Under the so-called large-detuning limit, i.e., $\delta\gg
g,\Omega_{0}$, the second-order contribution to $U_{I}(t)$ is
significantly important than the first order one. This is because
that the former involves terms linear in time, whereas the latter
involves the terms that are just oscillatory or constant in time.
Therefore, we can only retain the above second-order terms and
rewrite the above time-evolution operator (14) as
\begin{equation}
U_{I}(t)\approx 1-\left\{\frac{\Omega _{L}^{2}}{\delta }\sigma _{22}+\frac{%
g^{2}}{\delta }aa^{\dag}\sigma _{11}+\frac{\Omega _{0}g}{\delta
}[\sigma _{21}ae^{i\phi}+\sigma
_{12}a^{\dag}e^{-i\phi}]\right\}t=1-iH_{\rm eff}t,
\end{equation}
with an effective Hamiltonian
\begin{equation}
H_{\rm eff}=\frac{\Omega _{0}^{2}}{\delta }\sigma
_{22}+\frac{g^{2}}{\delta }aa^{\dag}\sigma _{11}+\lambda_1
\left[\sigma _{21}ae^{i\phi}+\sigma
_{12}a^{\dag}e^{-i\phi}\right],\,\lambda_1=\frac{\Omega
_{0}g}{\delta}.
\end{equation}
Obviously, this effective Hamiltonian is equivalent (apart from the
unimportant Stark shifts) to the above $\phi$-dependent Hamiltonian
(5) with $m=1$. Therefore, after passing the cavity vacuum wherein
the driving parameter $\phi$ changes from $0$ to $2\pi$, the atom
undergoes the following evolution
\begin{eqnarray}
|\Psi_1\rangle \longrightarrow |\Psi_2(\tau_{1})\rangle
=e^{i\gamma_{01}+i\xi}\cos(\frac{\Omega_{R1}\tau_1}{2})
|2\rangle+i\sin(\frac{\Omega_{R1}\tau_1}{2})e^{-i\xi} |1\rangle.
\end{eqnarray}
\begin{figure}[tbp]
\includegraphics[width=9cm,height=6.6cm]{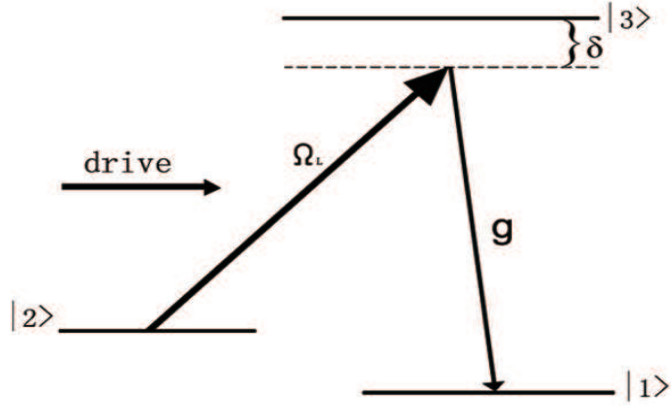}\newline
\caption{Schematic diagram of an three-level atom interacting with a
quantized field in $C$ cavity (see Fig.~2), which induces the
transitions $|3\rangle \leftrightarrow |1\rangle$ with Rabi
frequency $g$. In addition, a classical laser field driving the
transition $|3\rangle \leftrightarrow |2\rangle$ (with Rabi
frequency $\Omega_L=\Omega_0e^{i\phi}$) is applied to produce the
desirable $\phi$-dependent single-mode JCM.}
\end{figure}
\begin{figure}[tbp]
\includegraphics[width=13.5cm,height=8.6cm]{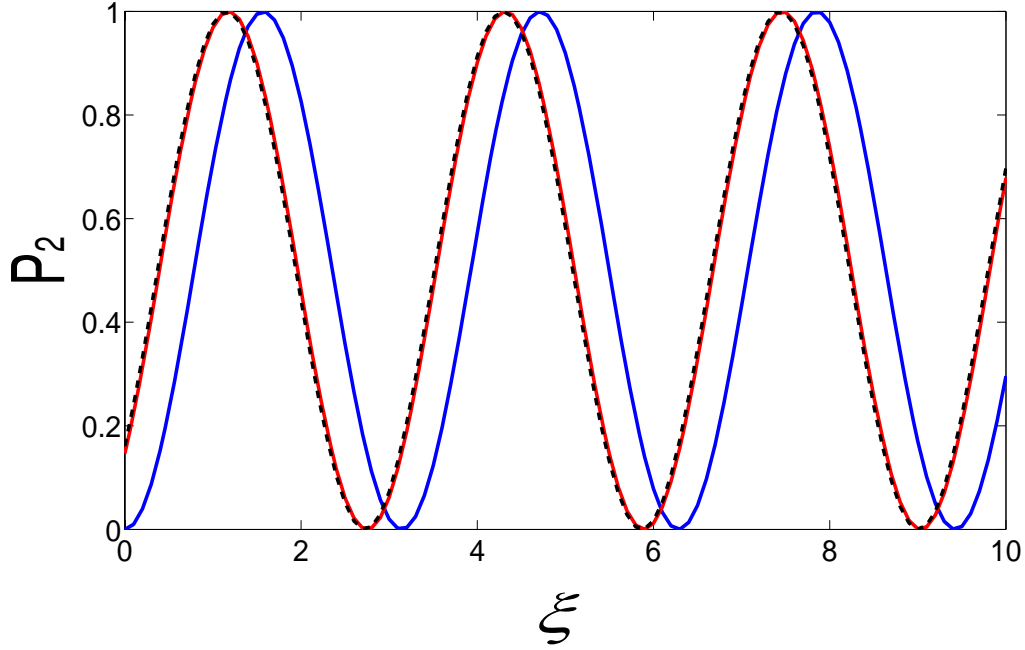}\newline
\caption{Experimental predictions to observe the vacuum-induced
Berry phase by measuring $P_2$, the probability of the atom being
detected in the state $|2\rangle$. This probability is a function of
the controllable parameter $\xi$ related to the atom interacting
with the cavity. Here, the blue curve corresponds to the Ramsey
interferometry without Berry phase, while the red curve shows the
situation in which a geometric phase shift ($\pi/4$) is induced.
Additionally, the black (dashed) curve shows that in the presence of
the cavity decay $\Gamma=1$ KHz, the $P_2$ is little influenced.}
\end{figure}
Here, $\gamma_{01}=\Omega_1/4$ is the vacuum-induced Berry phase
acquired in the pass of the cavity, and $\xi=\lambda_1\tau$ with
$\tau$ being the duration of the atom stayed in the cavity.

Furthermore, after passed the second Ramsey zone $R_2$, the atom
evolves to the state
\begin{eqnarray}
|\Psi_2\rangle \longrightarrow |\Psi_3(\tau_1,\tau_2)\rangle
=c_1(\tau_1,\tau_2)|1\rangle+c_2(\tau_1,\tau_2)|2\rangle,
\end{eqnarray}
with
$$
\left\{
\begin{array}{cc}
c_1(\tau_1,\tau_2)=e^{-i\xi}i\sin(\frac{\Omega_{R1}\tau_1}{2})\cos(\frac{\Omega_{R2}\tau_2}{2})+
e^{i\gamma_{01}+i\xi}i\cos(\frac{\Omega_{R1}\tau_1}{2})\sin(\frac{\Omega_{R2}\tau_2}{2}),\\
\\
c_2(\tau_1,\tau_2)=e^{i\gamma_{01}+i\xi}\cos(\frac{\Omega_{R2}\tau_2}{2})\cos(\frac{\Omega_{R1}\tau_1}{2})
-e^{-i\xi}\sin(\frac{\Omega_{R2}\tau_2}{2})\sin(\frac{\Omega_{R1}\tau_1}{2}).
\end{array}
\right.
$$
If the two Ramsey zones are properly set such that the condition
$\Omega_{R1}\tau_1=\Omega_{R2}\tau_2=\pi/2$ is exactly satisfied,
then the probability of detecting the atom in its upper level
$|2\rangle$ in $I$ is
\begin{equation}
P_2=|c_2(\tau_1,\tau_2)|^2=\frac{1-\cos(\gamma_{01}+2\xi)}{2}.
\end{equation}
If $\xi=n\pi$ is set inside the cavity, the above probability can be
further simplified to
\begin{equation}
\tilde{P}_2=\frac{1-\cos(\gamma_{01})}{2},
\end{equation}
which is directly related to the Berry phase acquired by the atom
flying across the high-Q quantized cavity $C$. Therefore, Berry
phase generated in the one-photon JCM could be observed by the above
Ramsey interference method.

Experimentally, the above one-photon Rabi frequency is set as
$g/2\pi\simeq50$kHz~\cite{14,15}. This implies that, if the solid
angle is required as $\Omega_{1}=\pi$, then the parameter
$\theta_{01}$ should be set to satisfy the condition
$\cos2\theta_{01}=1/2$. Since the parameter $\theta_{01}$ is
determined above by
$\theta_{01}=\arccos[\Delta_1/\sqrt{\Delta_1^2+4\lambda_1^2}]$, with
$\Delta_1=(\Omega_0^2-g^2)/\delta,\,\lambda_1=\Omega_0g/\delta$, the
Rabi frequency of the applied classical driving should be designed
as $\Omega_{0}/2\pi\simeq173kHz$.
On the other hands, in order to satisfy the large detuning condition
required above, i.e., $\delta\gg g,\Omega_{0}$, we may typically set
$\delta=3\Omega_{0}$ yielding $\lambda_{1}/2\pi\simeq15kHz$. This
means that the atom-field interaction cam perform $10$ complete Rabi
cycle during an effective atom-cavity interaction time of $0.6$ ms
~\cite{8}. This interaction time is manifestly shorter than the
decaying time ($1$ms) of the cavity (see, e.g.,~\cite{14}).

We now discuss how the dissipation of the cavity influence on the
observable effect of the vacuum-induced geometric phase. Following
the ref.~\cite{16}, the effective Hamiltonian of the atom-cavity
system becomes: $\tilde{H_{eff}}=H-i\Gamma\hat{n}/2$. With the same
procedure we can prove that, through a cyclic and adiabatic
evolution the acquired geometric phase reads
\begin{equation}
\gamma_{01}^{d}=\frac{\pi}{2}(1-Re\frac{(\Delta_{1}-i\Gamma/2)^{2}-4\lambda_{1}^{2}}
{(\Delta_{1}-i\Gamma/2)^{2}+4\lambda_{1}^{2}}).
\end{equation}
Since $\Gamma/R$ should be a perturbation quantity, we can expand
the above geometric to the second order in $\Gamma/R,
R=\sqrt{\Delta_{01}^{2}+4\lambda_{1}^{2}}$ and then obtain
\begin{equation}
\gamma_{01}^{d}\approx\gamma_{01}+\frac{\pi}{4}
\frac{\cos^{2}\theta_{01}}{8\sin^{2}\theta_{01}+16\sin^{4}\theta_{01}
+\cos^{4}\theta_{01}}(\frac{\Gamma}{R})^{2}.
\end{equation}
Consequently, the probability of the atom being detected in the
state $|2\rangle$ is changed to be
$\tilde{P}_2^{d}=[1-\cos(\gamma_{01}^{d})]/2$. Notice that in the
case of low decoherence, the lowest order correction of the expected
geometric phase is only quadratic in $\Gamma/R$, suggesting that the
field decoherence may not play such an important role in the
proposed experiment. This can be numerically verified from the
comparison in the Fig.~4: after considering the presence of a
typical cavity dissipation $\Gamma=1\mathrm{KHz}$~\cite{17}, the
probability of the atom being detected in the state $|2\rangle$ (the
black line) is almost unchanged. This means that the experimental
detection of the vacuum-induced Berry phase in JCM with the above
Ramsey interference is feasible, even in presence of the cavity
losses.

In summary, we have calculated the Berry phase of $m$-quantum JCM
and proposed an experimental setup to observe and measure such a
geometric phase induced by the vacuum field in an one-photon
single-mode JCM. Basically, geometric phases acquired by the
atom-field system are dependent of the number of photons in the
field. This is different from those attained in semi-classical
counterpart.
Our results show also that, for a common
$\bigtriangleup_{m}/\lambda_m\equiv\Delta/\lambda$, the more quantum
$m$ corresponds to the larger vacuum-induced Berry phase.
A Ramsey interference experiment with cavity QED is designed to
detect the vacuum-induced Berry phase.

\end{CJK*}

\section*{Acknowledgments}
This work was supported in part by the National Science Foundation
grant No. 10874142, 90921010, and the National Fundamental Research
Program of China through Grant No. 2010CB92304, and the Fundamental
Research Funds for the Central Universities No. SWJTU09CX078.

\end{document}